\def \ck           {\checkmark}

\documentclass[manuscript]{aastex}
\usepackage{times}
\usepackage{textcomp}
\usepackage{wasysym}
\usepackage{graphicx}
\usepackage{epstopdf}
\newcommand{\refs}{\par\noindent\hangindent=1pc\hangafter=1}
\begin{document}

\title{\textbf{\large Evidence for Two Populations of Classical Transneptunian Objects: \\ The Strong Inclination Dependence of Classical Binaries}}

\author {\textbf{Keith S.~Noll}}
\affil{\small\em Space Telescope Science Institute, 3700 San Martin Dr., Baltimore, MD 21218}
\email{\small noll@stsci.edu}

\author {\textbf{ William M.~Grundy}}
\affil{\small\em Lowell Observatory, 1400 W.~Mars Hill Rd., Flagstaff, AZ 86001}

\author {\textbf{ Denise C.~Stephens}}
\affil{\small\em Brigham Young University, Dept. of Physics and Astronomy, N145 ESC, Provo, UT 84062}

\author{\textbf{Harold F.~Levison}}
\affil{\small\em Dept.~of Space Studies, Southwest Research Institute, Ste.~400, 1050 Walnut St., Boulder, CO 80302}

\author {\textbf{ Susan D.~Kern}}
\affil{\small\em Space Telescope Science Institute}

\newpage

{\textbf{\hfil {Abstract\ \ \ } }}

\noindent{We have searched 101 Classical transneptunian objects for companions with the Hubble Space Telescope.  Of these, at least 21 are binary.  The heliocentric inclinations of the objects we observed range from 0.6-34$^\circ$.  We find a very strong anticorrelation of binaries with inclination.  Of the 58 targets that have inclinations of less than 5.5$^\circ$, 17 are binary, a binary fraction of 29$\pm{7\atop 6}$\%.  All 17 are similar-brightness systems.   On the contrary, only 4 of the 42 objects with inclinations greater than 5.5$^\circ$ have satellites and only 1 of these is a similar-brightness binary.  This striking dichotomy appears to agree with other indications that the low eccentricity, non-resonant Classical transneptunian objects include two overlapping populations with significantly different physical properties and dynamical histories.}


\smallskip

\noindent{ Running head: Classical Binaries Inclination Dependence}

\noindent{ Manuscript pages:  32}

\noindent{ Figures: 4}

\noindent{ Tables: 2}

\newpage

\section{\textbf{Introduction}}

Binaries are a well-established feature of the transneptunian population; more than 50 systems are known (Noll et al.~2008).  The number of transneptunian binaries (TNBs) is most usefully expressed as a fraction of the population; this fundamental quantity is among the first to be determined by large surveys.  Any such fraction must be qualified by the detection limits of the instrument configuration used to carry out the observations.  As the number of detected binaries increases, it is becoming clear that the fraction of detectable TNBs is a function of factors beyond instrument sensitivity (Stephens and Noll 2006).

One of the surprising facets of the transneptunian population is its dynamical complexity, a complexity that has fueled a proliferation of classification schemes.  While there is no generally agreed-upon system of classification ({\it e.g.} Elliot et al.~2005; Gladman et al.~2008; Lykawka and Mukai 2007), all of the proposed schemes recognize objects in stable, non-resonant, non-scattering, low-eccentricity ($e_\odot$ ${<\atop\sim}$ $0.24$) orbits as a distinct class.  These objects are generally referred to as ``Classical" transneptunian objects (TNO) and are mostly located between 42-48 AU.  The Classicals are sometimes further subdivided into two subgroups, Hot and Cold, based on their inclination.   The high-inclination Hot Classicals differ in measurable ways (color, absolute magnitude) from the low-inclination Cold Classicals and may have a distinct evolutionary history  (Tegler and Romanishin 2000; Levison and Stern 2001; Brown 2001).   As we show in this work, an additional distinction between these two subclasses is the fraction of objects that are binary.

We have been conducting a series of surveys of transneptunian objects using the Hubble Space Telescope since 2001.  In this paper we report on the result of searches for binary companions of 101 Classicals observed with the Near Infrared Camera Multi-Object Spectrometer (NICMOS) or the Advanced Camera for Surveys (ACS).  We find a remarkable concentration of binaries among the low inclination Classicals and, conversely, a marked absence of these systems among higher inclination Classicals.  The variation in the fraction of binaries with inclination appears to be another indicator that the Classicals are composed of two partially overlapping populations that may have originated in distinctly different parts of the early Solar System.

\section{Observations}

The data discussed in this paper come from observations made with Hubble Space Telescope (HST) from August 2002 through January 2006 plus two objects that we observed with HST in August 1998.   A complete list of the 101 Classicals we considered for this work can be found in Table 1. 

Most of the observations come from one of three HST snapshot programs, one using NICMOS and two using the ACS.  Targets in snapshot programs are selected randomly from a longer list of potential targets.  We managed the list to maintain a balance of dynamical types, but could not control which objects were observed.  All targets were required to have sufficiently well-known orbits that the uncertainty in their positions by the end of the observing cycle were ${<\atop\sim}$ 5 arcsec.

We identified targets in our snapshot programs by their motion relative to other sources in the field.  In order to ensure that targets would have sufficient relative motion we used the targeting windows in the HST phase 2 proposals to exclude times when the target motion was $<$ 0.18 arcsec/hr.  This eliminates only a negligible portion of the scheduling window for targets.

\subsection{NICMOS Observations}

Somewhat less than half of the Classicals in our sample were observed with NICMOS as part of HST programs 9060 (August 2002 - June 2003) and 7858 (August 1998).  The observations were made with the NIC2 camera, one of three separate cameras that are part of NICMOS.  We used the F110W and F160W filters corresponding, approximately, to the infrared $J$ and $H$ bands.  The NIC2 camera has a nominal pixel scale of 0.0759$\times$0.0754 arcsec/pixel and low geometric distortion.  At the NIC2 pixel scale, diffraction-limited HST images are critically sampled in the F160W filter, and are undersampled in the F110W filter.  Each target was observed at two dither positions.  At each dither position we obtained one image in each filter.  The observations are fully described by Stephens and Noll (2006).  In this earlier work, we detailed the techniques we used to identify a total of 9 binaries in NICMOS images.  For 3 of the 9 binaries we were able to identify the secondary easily via visual inspection.  All of these 3 were sufficiently well-separated that no modeling was required.  The remaining 6 binaries had smaller separations and were identified using a binary PSF fitting program.  Several of the binaries found from PSF fitting were subsequently confirmed using higher resolution HRC images including two Classicals, 1999 OJ$_4$ and (79360) 1997 CS$_{29}$.

\subsection{HRC Observations}
More than half the objects were observed using the High Resolution Camera (HRC), a component of the ACS that is currently inoperable.  The HRC has an anamorphic pixel scale of 0.0284$\times$0.0248~arcsec/pixel.  The HRC slightly oversamples the diffraction limited HST PSF which has a full-width-half-maximum (FWHM) of approximately 0.07 arcsec for the filters we used.

Most of the HRC data come from one of two identical HST Snapshot programs (10514 and 10800) that we carried out from July 2004 through January 2007.  The data taken as part of these two programs were obtained with the Clear filters in the HRC.  The effective bandpass of the Clear filter mode is determined by the wavelength response of the CCD detector; it has a FWHM of 520 nm centered at 620 nm.  This non-standard filter configuration of the HRC provided the maximum possible throughput at the cost of an approximate 10\% increase in the  FWHM of the point-spread function (PSF) compared to the diffraction limited FWHM.  The properties of the Clear filters are described in more detail by Gilliland and Hartig (2003).    

Most objects from programs 10514 and 10800 were observed with a sequence of four dithered exposures using the standard HRC box-dither pattern.  The exposure times were 300 seconds duration per dither position for most of these objects.  For 4 bright targets we used shorter 240 sec exposures at each position in the box dither pattern.   The objects that we imaged through December 2005 were observed with three dithered exposures of 500 seconds each.  With only three exposures, combined images typically have several artifacts resulting from triple-incident cosmic rays.  Because these can mimic faint companions, we changed the observing sequence to the four-exposure pattern starting in January 2006.  This pattern effectively eliminates all artifacts from multiply-coincident cosmic rays.  We note, however, that none of the three-exposure objects was found to have a faint companion and therefore there is no bias in the results as a result of these different patterns.  

For the sake of completeness, we also included in our analysis 7 bright targets observed by other programs; all have $H_{\rm V} \le 4.5$.  The Classical (20000) Varuna was observed by S.~Sheppard and K.~Noll in proposal 10555 with the HRC and was found to be single.  The observations used the F606W filter and were dithered in a four-point box-dither pattern with 620 sec exposures at each position.  Brown and Suer (2007) reported the detection of small satellites orbiting (50000) Quaoar and (55637) 2002 UX$_{25}$ as part of HST program 10545.  Brown and colleagues also observed 5 other apparently single Classicals including repeat observations of (20000) Varuna and (19308) 1996 TO$_{66}$.   The objects observed in 10545 were imaged with two 300 sec exposures and were not dithered.

\section{Analysis}

The results of our binary search are shown in Table 1 and are illustrated in Figures 1 and 2.  There is an obvious concentration of binaries at low inclination with a nearly complete absence of binaries at higher inclinations.  To quantify the anti-correlation between Classical binaries and inclination seen in Figures 1 and 2, we must first consider which objects to include in the analysis. 
The number of Classicals in our sample, and the statistical significance of our results, vary somewhat depending on which definition of Classical is used.  Ultimately, however, as we show in Table 2, the binary-inclination anticorrelation is significant regardless of how Classicals are defined.  

The homogeneity of the observations used to search for binaries is another important issue that must be addressed.  In order to increase the sample size we have included data sets that are not strictly homogeneous.  However, as we show below, the detection limits of these data sets are comparable and remaining biases can be addressed with some confidence.

\subsection{Definition of Classical}

The dynamical classification of transneptunian objects is an evolving and, to some degree, arbitrary subject.  All of the schemes recognize the Classicals, {\it i.e.} the non-resonant, low-eccentricity, non-scattering population, as a distinct dynamical class.  There are, however, disagreements about how to define the boundary between the Classicals and other dynamical classes.  In particular, there is disagreement about how far in inclination the Classicals should extend and how to define the boundary with the Scattered Disk.  The Deep Ecliptic Survey (DES) classification system (Elliot et al.~2005) defines Classicals by requiring that the objects be non-resonant and have a Tisserand parameter, $T_N$, relative to Neptune of $T_N > 3$ and a low eccentricity, $e < 0.2$.  The Tisserand parameter is defined as $T_N = a_N/a\ + 2[(1-e^2)a/a_N]^{1/2}\ cos(i)$ where $a$,$e$, and $i$ are the heliocentric semimajor axis, eccentricity, and inclination of the TNO and $a_N$ is the semimajor axis of Neptune.  Under this definition an object on a circular orbit at 45 AU must have $i < 18^\circ$ in order to be considered a Classical.  This definition excludes the low-eccentricity objects in this region that can have inclinations as high as $\sim 30^\circ$ in our sample.

Gladman et al.~(2008; GMV) propose an alternative definition where the Tisserand parameter is not used.  Instead, Classicals are the objects that are not resonant and are not actively being scattered by Neptune (in 10 Myr orbit integrations).  GMV also differ in that they employ a slightly higher eccentricity cutoff than does the DES, $e < 0.24$, for inclusion in the Classical group.  Other authors ({\it e.g.} Lykawka and Mukai 2005) sometimes limit Classicals to semimajor axes between the 2:3 and 1:2 resonances (39.4 - 47.7 AU) where most Classicals occur, but neither the DES nor the GMV schemes do so.  

In this work we have opted to work with the broadest definition of Classicals, {\it i.e.} the GMV definition.  The most prominent difference between the DES and GMV definitions is that the GMV Classicals include a number of objects at higher inclinations that are classified as Scattered-Near by the DES.  In Table 1 and Figures 1 and 2 we identify those objects that are not Classicals by the DES definition.  Most of these objects are classified as Scattered-Near in the DES system.  It is important to note that the conclusions of our work do {\bf not} depend on which classification scheme is used.  The step-function-like decrease in binaries at $i \approx 5.5^\circ$ is robust with any of the existing definitions of Classicals.

For convenience, we have also adopted the terms Cold Classicals and Hot Classicals to refer to Classicals with inclinations below and above some cutoff, respectively.  The value of the cutoff is not predetermined, but generally is near 5$^{\circ}$.  Considering only binaries, we would put the boundary slightly higher, at 5.5$^{\circ}$, but we consider this difference to be small.   

Finally, we note that throughout this paper when we refer to ``inclination", unless otherwise qualified, we are referring to the inclination relative to the invariable plane.  Because inclination can change over time, we use the mean inclination calculated from 10 Myr integrations (Elliot et al.~2005).  We list both the osculating heliocentric inclination, $i_\odot$, and the 10 Myr mean inclination with respect to the invariable plane, $\bar i$, in Table 1.  The objects in Table 1 are listed in order of increasing $\bar i$.  As a comparison shows, the distinction between $\bar i$ and $i_\odot$ makes only minor differences and does not affect the conclusions of this paper.  

\subsection{Detection Limits}

Any discussion of binary statistics must include a quantitative estimate of detection limits.  One of the advantages of observing with HST is the inherent stability of the PSF.  Observations made with the same HST instrument, filter and exposure will have essentially identical detection limits.  Small variations in focus and jitter account for generally negligible differences between individual exposures.  However, even in the ideal case of observations made with the same HST instrument and observing sequence, detection limits still depend on the brightness of the primary and the separation of the components in a non-trivial way that makes it impossible to state a single detection limit, particularly for close companions.  The data included in this paper were mostly obtained with one of two observational configurations.  We discuss the detection limits for each of these configurations separately below.

Detection limits for the observations made with NICMOS were discussed in detail by Stephens and Noll (2006).    At large separations from the primary, the detection limit is determined by the filter and exposure time and, to a lesser extent, the background from cosmic ray residuals and scattered earthshine.  According to the NICMOS Exposure Time Calculator (Arribas et al.~2004), the 5$\sigma$ detection limit for a single 512 sec exposure with the F160W filter occurs for an H-band magnitude of 22.6 (using aperture photometry; PSF fitting can improve the S/N by roughly a factor of 2).  The median magnitude of the objects observed with NICMOS is $H_{\rm med} = 21.15$.   In tests with embedded artificial secondaries separated by at least one full resolution element, Stephens and Noll found half of objects with $H$=23.5 consistent with the predicted S/N of $\sim$2 for this magnitude.   

For separations of less than $\sim$2 pixels, the local background for a possible secondary is dominated by the PSF of the primary.  The detection limit in these cases is determined, in part, by the brightness of the primary, as shown by Stephens and Noll (2006) in their Figure 2.  It is interesting to note that given sufficient S/N, i.e. bright sources, secondaries can be found with separations that are a fraction of a pixel.  However, detecting these objects at separations smaller than the Nyquist limit requires a binary PSF-fitting program like the one discussed by Stephens and Noll.  PSF-fitting was carried out for all of the NICMOS data and all but one of the detected NICMOS binaries was found at a separation of 0.061 arcsec (0.8 pixels) or greater.   All of the NICMOS binary detections were significant at 3$\sigma$ or better. 

Detection limits for our observations with the HRC similarly divide into two different regimes.  For objects far enough from other sources, the background is determined by sky brightness, dark current, and read noise.  The expected 5$\sigma$ detection limit for single 300 second integrations with the Clear filters is $V = 26.5$.  By combining all four images, the limit drops to $V = 27.2$ (Noll et al.~2008).  Analysis of observed backgrounds shows a very close match to the predicted performance.   The median magnitude of the objects in the HRC sample is $V_{\rm med} = 22.8$.  

At small separations the background for a potential secondary in HRC images is determined by the PSF of the primary.  As discussed for NICMOS above, the detection limit becomes a complex function of the brightness of the primary and the relative position of the secondary.  For the HRC this becomes even more complex because of the azimuthal asymmetry of the PSF.  We estimated the detection limits for close binaries with a series of implanted secondaries at a range of separations (Noll et al.~2006a, 2008).  To give one specific example, at a separation of 3 pixels ($\sim$0.075 arcsec), typical of the close binaries found with the HRC, the detection limit is approximately $V = 25$ for a primary with $V = 23$.   

While combining data from different instruments with differing detection limits is not ideal, it does give us the advantage of increasing the sample size.  Because PSF-fitting for binary identification has been completed for the NIC2 data, but not for the HRC data, the relative separation limits for the two data sets are approximately equal, roughly 0.06-0.07 arcsec.   The two instruments differ substantially in their ability to detect faint secondaries.  For widely separated binaries the HRC can go significantly deeper relative to the primary ($\Delta_{\rm mag}\sim$4 vs.~$\Delta_{\rm mag}\sim$2 magnitudes) for a ``typical'' target.  A comparable advantage for the HRC holds at smaller separations.  However, as shown by Noll et al.~(2006a, 2008), the magnitude difference for Classical binaries, for both NICMOS and HRC, is clustered at $\Delta_{\rm mag} < 1$.  This clustering suggests that the shallower NICMOS images have missed few secondaries.   As can be seen in Figures 1 and 2, the trend of binaries with inclination is clearly present in both the NICMOS-only and HRC-only halves of our data set.  Thus we conclude that combining the two data sets serves our purpose of improving the statistics without introducing unnecessary bias.

We mention one caveat with regard to our list of detected binaries.  PSF fitting has not been carried out for the HRC data reported here.  When PSF fitting is completed it is possible that additional close binaries will be detected at separations of less than 0.06 arcsec.  However, the NICMOS and HRC data sets will then not be directly comparable because the number of binaries increases rapidly at small separations (Noll et al.~2008).  Thus, the current situation where both NICMOS and HRC angular resolution limits are similar, is optimal for the purposes of this paper.

\subsection{Statistics}

We applied two statistical tests to our samples in order to quantify the significance of the binary fraction as a function of inclination.  The results are summarized in Table 2.  

We used the Kolmogorov-Smirnov (K-S) test (Press et al.~1992) to assess the likelihood of the observed distribution of binaries with inclination arising by chance from a single population with a fixed fraction of binaries.  For this test we divide the sample into singles and binaries and for each group create an ordered list in inclination.  The result of this ordering is plotted in Figure 3 where the cumulative fraction of objects in each list is plotted as a function of inclination.  The K-S test makes use of the so-called $D$-statistic, which is the maximum vertical distance between the two cumulative distribution functions and is invariant with respect to the scaling of the abscissa.  Because we do not have an {\it a priori} expectation for the distribution of either singles of binaries with inclination, we use the two-sample K-S test to convert the $D$-statistic into a probability, $p_{\rm KS}$.  As shown in Table 2, the probability that the null hypothesis is correct, {\it i.e.} that the binaries and singles have the same distribution in inclination, is $p_{\rm KS}$ = 0.8 - 6.4\% depending on which sample is used.  The largest uniform sample, GMV Classicals with $H_{\rm V} > 5$, gives a 99.2\% probability ({\it i.e.} $1-p_{\rm KS}$) that singles and binaries have different inclination distributions.

A size-inclination correlation among Classicals, like the one proposed by Levison and Stern (2001), would bias our results if there were a change in the rate or type of binaries with the size of the object.   Just such a change has been suggested by Brown et al.~(2006) who find that a high fraction of the largest TNOs have small satellites, possibly the result of collisions.  The smaller Classicals are dominated by similar-sized binaries that most plausibly formed from capture (Lee et al.~2007, Noll et al.~2008).  In order to account for the observed changes in the inclination distribution and in the type of binaries that are detectable as a function of absolute magnitude, $H_{\rm V}$ (an observable proxy for size), we have considered samples with and without a cutoff in $H_{\rm V}$. 

We checked the sensitivity of the K-S test to the choice of $H_{\rm V}$ cutoff by iteratively computing $p_{\rm KS}$ while allowing the $H_{\rm V}$ cutoff to vary.  In all cases we considered only an $H_{\rm V}$ minimum cutoff ({\it i.e.} a maximum object size).  The results for both GMV and DES Classicals are plotted in Figure 4.  A broad minimum in the probability occurs for a cutoff at $H_{\rm V} \approx 5$.  This is not surprising given the lack of large objects at low inclinations and the prevalence of smaller satellites around the larger objects which led us to consider a cutoff in the first place.  

It is illuminating to compute the fraction of binaries in the different subgroups we have defined.  Using the GMV sample we find that 17 of 58 Classicals with $i < 5.5^{\circ}$ are binary.  Expressed as a percentage, this is 29.3$\pm {7.2\atop6.3}$\% (1$\sigma$).  If we limit this sample to objects with $H_{\rm V}\geq 5$, we eliminate only 1 object, a single.  The ratio is then 17 of 57 for a binary fraction of 29.8$\pm {7.3\atop6.4}$\%.  For higher inclination objects, $i >5.5^{\circ}$, the GMV sample includes 43 objects, 4 of which are binaries.  This corresponds to a binary fraction of 9.3$\pm {6.7\atop4.4}$\%.  Eliminating the brightest objects, $H_{\rm V} < 5$, from this sample reduces the total to 34 objects with only a single binary.  The binary fraction of the Hot Classicals is then only 2.9$\pm {6.5\atop2.4}$\%. 

We also conducted a $\chi^2$ test on our data after binning it into 2$\times$2 contingency tables (Table 2).  We use Yates' correction for continuity which is necessary for samples where small sample sizes occur in one or more of the cells (Preacher 2001).  In order to carry out this type of test it is necessary choose an inclination boundary for binning.  We chose an inclination of $\bar i$=5.5$^\circ$ which maximizes the probability that the two groups defined in this way differ.  The test could be conducted for other inclination cutoffs, if desired.  An inclination boundary of $\bar i$=4.6$^\circ$ (Peixinho et al.~2004, Gulbis et al.~2006) yields a result that is only slightly less significant than we found for a boundary at $\bar i$=5.5$^\circ$.

\section{Discussion}

Why should binaries be frequent around Classical objects at low inclination but absent around Classicals at higher inclinations?  From a strictly logical point of view, there are only two possibilities.  Either the high inclination population did not form many binaries in the first place or it did but subsequently lost them.  We describe variants of each of these possible scenarios below and consider their implications.

\subsection{Environmental Effects}

In an early attempt to understand the remarkable color diversity of the Kuiper Belt, Stern (2002) found an apparent $\approx3\sigma$ correlation between the ``rms collision velocity'', $V_{\rm rms}$, and color.  In this formulation $V_{\rm rms}$, defined as $V_{\rm rms} \propto (e^2 + i^2)^{1/2}$, provided a possible link between physical properties and inclination.  However, this model was critiqued by Th\'ebault and Doressoundiram (2003) who pointed out that $V_{\rm rms}$ is indicative of excitation for a particular body, but is not directly related to impact energy, which must take into account the full dynamical structure of the Kuiper Belt.  Th\'ebault and Doressoundiram modeled impact energy for several possible impactor distributions and found that impact energy correlations, when present at all, were more strongly correlated with eccentricity than with inclination.  Because of the lack of correlation with inclination, Th\'ebault and Doressoundiram rejected the impact energy hypothesis as a possible source of the observed color-inclination correlation in the Kuiper Belt.  We make the same conclusion with regard to the inclination-dependence of binaries that we have found.

We are unaware of any other environmental mechanism that can produce variations in physical properties with inclination, {\it in situ}.  Thus, we conclude, the inclination dependence of binaries cannot have evolved from a single homogeneous precursor population.  

\subsection{Two Populations} 

The logical alternative explanation for the observed inclination-binary dichotomy is that the objects currently defined as Classicals are, in fact, two different populations that partially overlap in orbital parameter space with similar distributions in $a$ and $e$, but different $i$ distributions.  This suggestion has been made before based both on observational evidence and on theoretical grounds.  We briefly summarize each line of evidence below.

\noindent{\it\underbar{Ensemble inclination}\ \ }  Brown (2001) showed that the observed inclination distribution of 379 TNOs could be best modeled as a combination of two simple functions (Gaussians) distinguished by different FWHM.  This result has recently been confirmed with an independent data set of DES-discovered TNOs  (Elliot et al.~2005).  Both Brown's and Elliot et al.'s samples are heavily weighted towards Classicals.  Brown carried out a separate calculation for just the 251 Classicals (defined as $a > 40 AU$ with an unspecified limit on $e$) in his sample and fit an inclination distribution with two gaussians of $\sigma$ of 2.2$^\circ$ and 17$^\circ$.  

The bimodality of the inclination distribution strongly suggests, but does not prove, the existence of two independent populations.  Brown (2001) argues that a Gaussian is the natural distribution for the ecliptic inclination distribution of a population of randomly perturbed zero-inclination precursors, which, if it applies in this case, would require two independent populations.  On the other hand, Brown also points out that a bimodal distribution similar to the one he finds for Classicals could result from a close stellar encounter.  To definitively identify two genetically distinct populations there must be measurable physical differences between the two components of the inclination distribution that cannot be attributed to evolution of an initially homogeneous precursor population.

\noindent{\it\underbar{Absolute magnitudes}\ \ }  Levison and Stern (2001) noticed a correlation between size (using absolute magnitude as a proxy) and inclination among Classicals.  Objects with low inclination are lacking among the largest known Classicals but are abundant at smaller sizes.  Levison and Stern proposed that such a situation could arise if Classicals were made up of two overlapping populations: a dynamically hot population with a larger maximum object size and higher maximum inclination, $i_{\rm max}\approx$ 35$^\circ$, and a dynamically cold population with smaller objects and inclinations of less than $i_{\rm max} \approx$ 5$^\circ$.  Even though many more objects are now known, including larger objects at all inclinations, the lack of low-inclination objects among the largest Classicals that prompted their work has persisted. 

\noindent{\it\underbar{Color and Albedo}\ \ }  The colors (typically B-R or B-V) of Classicals are also observed to vary with inclination (Tegler and Romanishin 2000, Trujillo and Brown 2002, Doressoundiram et al.~2002, Tegler et al.~2003, Peixinho et al.~2004, Gulbis et al.~2006) with low inclination Classicals ($i < 4.5^\circ$) appearing to be almost uniformly red, while higher inclination Classicals ($i > 4.5^\circ$) include both red and neutral colors.  


Grundy et al. (2005) noted a possible trend in albedo and inclination with low inclination Classicals having a higher albedo than high inclination Classicals, albeit with a small sample size.  Brucker et al. (2007) have now confirmed this trend with a larger sample.  

Both the color and albedo trends are consistent with the hypothesis that the Classicals include two populations with distinct surface properties.  As was true for binary frequency, the lack of a physical mechanism correlated to inclination argues against an {\it in situ} evolution of the observed color-inclination and albedo-inclination correlations.

\noindent{\it\underbar{Formation models}\ \ }  Theoretical models and numerical simulations are often used to examine the processes and initial conditions that could have produced the observed dynamical features of the Kuiper Belt.   A common thread running through all of the models discussed below is that some or all of the objects in the Kuiper Belt originated elsewhere in the protoplanetary disk.  This leads to a natural, but unquantified, source for physical differences, as we discuss below.  

Gomes (2003) reported the results of numerical simulations of planetesimal scattering by Neptune as it migrated outward.  The simulations were able to reproduce the high inclination populations of the Kuiper Belt, including the Hot Classicals, from a small fraction of Neptune-scattered planetesimals that are temporarily trapped in resonances.   In this scenario, the  high inclination objects formed closer to the Sun than the lower inclination Cold Classicals that are a remnant of an outer disk that was not strongly disturbed during planet migration.  The different origins can account, qualitatively, for the observed differences between Hot and Cold Classicals (size, color, albedo, and binaries).  A testable corollary of the Gomes model is that all of the high inclination populations, Scattered, Resonant, and Hot Classicals, should have similar rates of binaries.   The statistics on these populations are, however, currently insufficient.

Levison and Morbidelli (2003) proposed an alternative mechanism that could account for the Cold Classicals.  As a possible source for low inclination objects in the Kuiper Belt they examined the effects of the 1:2 resonance as it migrated outward into a planetesimal disk ahead of Neptune.  They found that leakage of objects temporarily trapped in the 1:2 resonance could approximately reproduce the dynamically cold population between 40 and 48 AU under some circumstances.  When combined with Gomes (2003), this model implies that both Hot and Cold Classicals formed closer to the Sun than their current location, but nevertheless with the Cold Classicals at larger heliocentric distances.  Thus, the qualitative basis for physical differences is preserved.

The Nice model (Levison et al.~2007a) is a comprehensive model that provides a framework for understanding the dynamical properties of the giant planets, the Kuiper Belt, Jupiter and Neptune Trojans, and the Late Heavy Bombardment.  In the Nice model, all of the objects in the current Kuiper Belt were scattered from smaller heliocentric distances during a period a giant planet orbital instability and migration.  Levison et al.~find a tendency for survivors scattered from the inner disk to have larger inclinations, on average, than objects that originated further out in the disk.   Interestingly, they also find that objects from the inner disk have more encounters with Neptune before arriving in the Kuiper Belt than do objects that start out at larger heliocentric distances, thus providing a possible mechanism for differences in binary survival.  

When considering differences in scattering history, it is worth keeping in mind that the disruption of binaries through scattering events has not yet been shown to be significant.  Two binaries that are currently in unstable Centaur-like giant-planet-crossing orbits are known (Noll et al.~2006b).  Orbit integrations suggest it is likely they will survive their $\sim 10^7$ year lifetime in these orbits which will include on the order of 100 close encounters with giant planets.  Whether or not the scattering events experienced by Centaurs in the present solar system are comparable in strength and number to the scattering events experienced by planetesimals in the early Solar System is unknown.

In a model meant to explain the presence of the 2003 EL$_{61}$ collision family (Barkume et al.~2006, Brown et al.~2007), Levison et al.~(2007b) explore the possibility that the Kuiper Belt's high inclination population may be composed of remnants of collisions between Scattered Disk objects.  Like the Gomes (2003) model, this mechanism can qualitatively explain difference between high and low inclination populations based on different origins.  In addition, it offers a novel explanation for a lack of binaries among high inclination objects that may have been formed by binary-disrupting collisions. 

At the moment, there is no way to discriminate between the various models that have been proposed that include a potentially separate origin for Hot and Cold Classicals.  It is possible that physical properties, including binary fraction, can be used to test for relationships among other subpopulations in the Kuiper Belt.  Any such relationships (or lack of relationship) might then be used to constrain, and possibly eliminate, some of the competing models of the early Solar System.

\subsection {Heterogeneity in the Protoplanetary Disk}

The contrast in physical properties between Cold and Hot Classicals stands out as the most distinct example of physical heterogeneity tied to dynamical status in the present-day Kuiper Belt.  Cold Classicals are tightly clustered at low inclinations, are deficient in large objects, are almost uniformly red, have high albedo, and a high fraction are binary.  Hot Classicals have a much broader inclination distribution, have a range of colors tending towards neutral, have lower albedo, and are almost uniformly single at HST detection limits.  The only viable explanation we (and others) have found for this diversity of properties, including the sharp cutoff in binaries, is that the Classicals are a heterogeneous group of objects with at least two distinct populations of objects having different origins and evolutionary histories.   

The two-population hypothesis leads to an interesting question:  How much of the current diversity is primordial and how much is the result of subsequent evolution?  Binaries may be particularly useful probes with respect to this question.  Binaries cannot form by dynamical capture at the volume density of the present-day Kuiper Belt; they must have formed in the much denser protoplanetary disk ({\it c.f.}~Noll et al.~2008 and references therein).   And, as we noted above, what analysis there is suggests that the relatively tightly bound binaries we find in the Kuiper Belt today are difficult to break up (Petit and Mousis 2004, Noll et al.~2006b).  If so, the variation in binaries that we see in the Classicals is the result of an equivalent variation that must have existed within the protoplanetary disk itself, with different populations mapped onto different locations in the protoplanetary disk.  From this, it is a relatively small step in logic to hypothesize that the other inclination-correlated physical properties of the Classicals also have their origin in a protoplanetary disk that is radially stratified.  What is more difficult to reconcile, perhaps, is the requirement in some of the current models that this stratification occur over a relatively small range of heliocentric distances.   A challenge, both for models of the early Solar System and for observations of extrasolar disks, is to understand on what scale and by what mechanism(s) variable physical properties in planetesimals can arise and persist in a dense circumstellar disk.

The existence of a TNO population with a very high incidence of binaries, such as the Cold Classicals, may also provide constraints on the amount of collisional evolution that has occurred since the binaries were formed.  Petit and Mousis (2004) argued that, under certain assumptions, a collisional environment capable of disrupting individual objects would have been even more effective at disrupting binaries.  Levison et al.~(2007c) noted this argument in their assessment of mechanisms capable of removing the much larger mass of planetesimals presumed to have been present during the accretional phase of the protoplanetary disk.  An interesting question is whether different populations of TNOs experienced different amounts of collisional evolution in the time since they left their nascent environment. The binary fraction might help us answer this question.

\section{Conclusions}

Binaries appear to be nearly absent among 100-km-sized Hot Classicals,  {\it i.e.}~low-eccentricity, non-resonant Kuiper Belt objects with absolute magnitudes of $H_{\rm V} \geq 5$ and inclinations of $i \geq 5.5^{\circ}$.  Only one such pair has been found, the binary 2001 QC$_{298}$ (which is classified as a Classical by only one of the two systems we considered).  Conversely, at inclinations of $i < 5.5^{\circ}$ binaries are common.  A large percentage, 29$\pm {7\atop 6}$ \%, of low-inclination Classicals have detectable binaries that are separated by $s > 0.06$ arcsec and with a magnitude difference of $\Delta_{\rm mag} < 2$. 

The largest Classicals, those with absolute magnitudes of $H_{\rm V} < 5$, are different.  Of the 11 objects that fit this description, only 1 has an inclination less than 5.5$^{\circ}$.  Satellites have been found around 3 of the 11.  However, all 3 systems have small secondaries compared to the primaries, $\Delta_{\rm mag} > 2$, unlike any of the more numerous, smaller Classical binaries.  One of these, (120347) 2004 SB$_{60}$ has heliocentric orbital elements that are similar to 2003 EL$_{61}$ and its possible collision family (Brown et al.~2007), although its neutral infrared colors (Kern, private communication 2007) make an association with this group unproven.  All of these factors suggest that the satellites of the largest Classicals may be collisional in origin as has been hypothesized for the largest TNOs generally (Brown et al.~2006).  While this leads to an unsatisfying complexity in origin of binaries among the Classicals, it may be an unavoidable byproduct of the fact that the Classicals are a composite population.  

Of the two logical possibilities for the origin of an inclination dependence for binaries, only the two-populations hypothesis appears to be consistent with available evidence.  Several of the scenarios for Kuiper Belt formation have natural explanations for the existence of these two populations.  In all these models, these two parent populations are formed at different locations in the protoplanetary disk.  We are then left with the unsolved question of how, exactly, binary fraction, color, size, and possibly other physical properties varied as a function of heliocentric distance in the early Solar System.  Because the same processes could apply in other circumstellar disks, further investigations into these questions have an intrinsic value that goes beyond our own local example of a planetary system.

\bigskip
\bigskip

\acknowledgments {This work is based on observations made with the NASA/ESA Hubble Space Telescope. These observations are associated with programs \#~9386, 10514, and 10800.  Support for these programs was provided by NASA through a grant from the Space Telescope Science Institute, which is operated by the Association of Universities for Research in Astronomy, Inc., under NASA contract NAS 5-26555.  HFL is grateful the PG\&G, OPR,  and Origins for continuing support.  We also thank S.~Sheppard and M.~Brown for permission to cite their unpublished work.}

\bigskip
\bigskip
\newpage

{\baselineskip=16pt\it

\noindent{{\bf Note added in proof:}\ \   During the preparation of this manuscript additional observations of TNOs were made with HST by the authors.  These additional observations were made using the WFPC2 in May and June 2007 following the on-orbit failure of the ACS/HRC.  The targets were observed on the PC chip (45 milliarcsec pixel scale) with an observing sequence of four dithered 260 second exposures through the F606W (wide-V) filter.  The PC is less sensitive to binary companions than the HRC because of the coarser pixel scale (slightly undersampled at 600nm), the larger overheads and consequent shorter integrations, and the lower sensitivity of WFPC2.  }

We observed a total of 31 Classicals in this period with mean inclinations relative to the invariable plane, $\bar i$, ranging from 0.94$^{\circ}$ to 12.48$^{\circ}$.  Most of these, 27 of 31, are Cold Classicals, {\it i.e.} they have $\bar i  <$ 5.5$^{\circ}$.  We found 5 binaries in our sample.  As with the ACS and NICMOS sample, the binaries have low inclinations.  The newly identified binaries are: 
(119067) 2001 KP$_{76}$ was observed on 2007 May 8 13:39 (UT) and found to be binary.  This object is found to be a classical by the DES team for the nominal orbital elements, although the 3-sigma range includes the possibility that it is in the 4:7 mean motion resonance with Neptune.  GMV find it to be in the 4:7 resonance.  The mean inclination for this object in 10 Myr integrations is $\bar i$ = 4.11$^{\circ}$.
2005 PR$_{21}$ was observed on 2007 May 10 12:22 (UT) and was observed to be a binary.
It is classified as a Classical both by the DES and by GMV.  It has a mean inclination of $\bar i$ = 2.69$^{\circ}$.
(160256) 2002 PD$_{149}$ was observed on 2007 May 22 04:30 (UT).  This object is a widely separated binary, s=0.74$\pm$0.01 arcsec, potentially detectable from the ground.  The mean inclination of this binary is $\bar i$ = 3.30$^{\circ}$.  
2001 QQ$_{322}$ was found to be binary from observations taken starting at 2007 June 15 00:28 (UT).  The mean inclination with respect to the invariable plane is $\bar i$ = 2.38$^{\circ}$.
(160091) 2000 OL$_{67}$, observed 2007 June 26 09:47 (UT) was found to be a binary.  ItÕs mean inclination is $\bar i$ = 3.48$^{\circ}$.

Objects observed with the PC that appeared to be single are listed with their mean inclination shown in parentheses.  The apparently single objects are: (85633) 1998 KR$_{65}$ (0.94$^{\circ}$), 1998 HP$_{151}$ (1.01$^{\circ}$), 2002 PO$_{149}$ (1.28$^{\circ}$), 2001 HZ$_{58}$ (1.34$^{\circ}$), 2001 OQ$_{108}$ (1.44$^{\circ}$), 2004 PX$_{107}$ (1.45$^{\circ}$), (24978) 1998 HJ$_{151}$ (1.45$^{\circ}$), 2000 ON$_{67}$ (1.67$^{\circ}$), (137294) 1999 RE$_{215}$ (1.72$^{\circ}$), (66452) 1999 OF$_{4}$ (1.80$^{\circ}$), 2003 QG$_{91}$ (1.81$^{\circ}$), 2003 QV$_{90}$ (1.96$^{\circ}$), 2001 KF$_{76}$ (2.20$^{\circ}$), 2001 QZ$_{297}$ (2.26$^{\circ}$), 2000 JF$_{81}$ (2.44$^{\circ}$), 2003 QA$_{92}$ (2.49$^{\circ}$), 2001 QX$_{297}$ (2.67$^{\circ}$), 2001 QD$_{298}$ (2.78$^{\circ}$), 2003 QF$_{113}$ (2.99$^{\circ}$), 2000 OH$_{67}$ (3.92$^{\circ}$), 2001 QB$_{298}$ (4.31$^{\circ}$), 2001 QR$_{297}$ (4.46$^{\circ}$), 2003 SP$_{317}$ (5.70$^{\circ}$), 2000 PX$_{29}$ (5.81$^{\circ}$), 2002 PD$_{155}$ (6.92$^{\circ}$), 2004 PZ$_{107}$ (12.48$^{\circ}$).

The addition of 5 new binaries found with the PC brings the total of new Classical binaries reported in this paper to 26.  15 of these are reported here for the first time.  The main conclusions of this paper are not altered by the inclusion of the PC data in the analysis.

}

\newpage

%
%

\centerline\textbf{ REFERENCES}
\bigskip
\parskip=0pt
{\small
\baselineskip=11pt

\refs Arribas, S., McLean, D., Busko, I., and Sosey, M.  2004.  New exposure time calculator for NICMOS (imaging): Features, testing and recommendations.   {\it Instrument Science Report NICMOS 2004-002}, STScI.
\\

\refs Barkume, K., Brown, M.E., Schaller, E.L. 2006.  Discovery of a collisional family in the Kuiper Belt.  Bull.~Amer.~Astron.~Soc., 38, \#44.06.
\\

\refs  Brown, M.E.  2001. The Inclination Distribution of the Kuiper Belt.  Astron.~J., 121, 2804-2814.
\\

\refs Brown, M.E. and Suer, T.A.  2007.  Satellites of 2003 AZ\_84, (50000), (55637), and (90482).  IAU Circ., 8812, 1.
\\

\refs Brown M.E.,  and 14 colleagues.  2006.  Satellites of the largest Kuiper Belt objects.  Astrophys.~J., 639, L43-L46. 
\\

\refs Brown, M.E., Barkume, K.M., Ragozzine, D., Schaller, E.L. 2007.  A collisional family of icy objects in the Kuiper Belt.  Nature, 446, 294-296.
\\

\refs Brucker, M., Grundy, W.M., Stansberry, J., Spencer, J., Buie, M., Chiang, E., Wasserman, L.H.  Determining the radii of sixteen transneptunian bodies through thermal modeling.  BAAS 39, 52.05 (abstract).
\\

\refs Doressoundiram, A., Peixinho, N.,  deBergh, C., Fornasier, S., Th\'ebault, P., Barucci, M.A., Veillet, C.  2002.  The color distribution in the Edgeworth-Kuiper Belt. Astrophys.~J., 124, 2279-2296.
\\

\refs Elliot, J.L., Kern, S.D., Clancy, K.B., Gulbis, A.A.S., Millis, R.L., Buie, M.W., Wasserman, L.H., Chiang, E.I., Jordan, A.B., Trilling, D.E., Meech, K.J.   2005.  The Deep Ecliptic Survey: A search for Kuiper Belt objects and Centaurs. II. Dynamical classification, the Kuiper Belt plane, and the core population. Astrophys.~J., 129, 1117-1162.
\\

\refs Gilliland, R. and Hartig. G.  2003. Stability and Accuracy of the HRC and WFC Shutters.  Instrument Science Report ACS 2003-03, STScI, Baltimore.
\\

\refs Gladman, B., Marsden, B.G., VanLaerhoven, C. 2008. Nomenclature in the outer Solar System.  In: Barucci, A., Boehnhardt, H., Cruikshank, D., Morbidelli, A. (Eds.) The Solar System Beyond Neptune, Univ. of Arizona Press, Tucson, pp.~43-57.
\\

\refs Gomes, R.S.  2003.  The origin of the Kuiper Belt high-inclination population.  Icarus, 161, 404-418.
\\

\refs Grundy, W.M., Noll, K.S., Stephens, D.C. 2005.  Diverse albedos of small trans-neptunian objects.  Icarus 176, 184-191.
\\

\refs Gulbis A.~A.~S., Elliot J.~L., Kane J.~F.  2006.  The color of the Kuiper Belt core.  Icarus, 183, 168-178.
\\

\refs Lee, E.A., Astakhov, S.A., Farrelly, D.  2007.  Production of trans-Neptunian binaries through chaos-assisted capture.  MNRAS, 379, 229-246.
\\

\refs Levison, H.F. and Stern, S.A.  2001.  On the size dependence of the inclination distribution of the main Kuiper Belt.  Astron.~J, 121, 1730-1735. 
\\

\refs Levison, H.F. and Morbidelli, A.  2003.  The formation of the Kuiper Belt by the outward transport of bodies during Neptune's migration.  Nature, 426, 419-421.
\\

\refs Levison, H.F., Morbidelli, A., Gomes, R., Tsiganis, K.  2007a.  Origin of the structure of the Kuiper Belt during a dynamical instability in the orbits of Uranus and Neptune.  Icarus, in press; ArXiv e-prints, 712, arXiv:0712.0553. \\

\refs Levison, H.F., Morbidelli, A., Vokrouhlick\'y, D., Bottke, W.F.   2007b.  On a scattered-disk origin for the 2003 EL$_{61}$ collisional family. Division of Dynamical Astronomy Meeting, 38, 10.02.  
\\

\refs Levison, H.F., Morbidelli, A., Gomes, R.,  Backman, D.  2007c.  Planet Migration in Planetesimal Disks.  {\it Protostars and Planets V}, B. Reipurth, D. Jewitt, and K. Keil (eds.), University of Arizona Press, Tucson, pp.~669-684.
\\

\refs Lykawka, P.S. and Mukai, T.  2005.  Long term dynamical evolution and classification of Classical TNOs.  Earth, Moon, and Planets, 97, 107-126.
\\

\refs Lykawka, P.S. and Mukai, T. 2007. Dynamical classification of trans-neptunian objects: Probing their origin, evolution, and interrelation.  Icarus, 189, 213-232.
\\

\refs  Noll, K.S., Grundy, W.M., Levison, H.F., Stephens, D.C.  2006a.  The relative sizes of Kuiper Belt binaries. Bull.~Amer.~Astron.~Soc., 38, \#34.03.
\\

\refs  Noll, K.S., Levison, H.F., Grundy, W.M., Stephens, D.C.  2006b.  Discovery of a binary Centaur.  Icarus, 184, 611-618.
\\

\refs Noll, K.S., Grundy, W.M., Chiang, E.I., Margot, J.-L., Kern, S.D.  2008.   Binaries in the Kuiper Belt.  In: Barucci, A., Boehnhardt, H., Cruikshank, D., Morbidelli, A. (Eds.) The Solar System Beyond Neptune, Univ. of Arizona Press, Tucson, pp.~345-363; astro-ph/0703134. 
\\

\refs Peixinho, N., Boehnhardt, H., Belskaya, I., Doressoundiram, A., Barucci, M.~A., Delsanti, A.  2004. ESO large program on Centaurs and TNOs.  Icarus, 170, 153-166.
\\

\refs Petit J.-M. and Mousis O. (2004) KBO binaries: how numerous were they? {\it Icarus, 168}, 409-419.
\\

\refs Preacher, K.J.  2001. Calculation for the chi-square test: An interactive calculation tool for chi-square tests of goodness of fit and independence [Computer software].  Available from http://www.quantpsy.org .

\refs Press, W.H., Teukolsky, S.A., Vetterling, W.T., Flannery, B.P.  1992.  Numerical Recipes in FORTRAN,  2nd edition.  Cambridge Univ.~Press, UK. 
\\


\refs Stephens, D.C. and Noll, K.S.  2006.  Detection of six trans-Neptunian binaries with NICMOS: A high fraction of binaries in the Cold Classical disk. Astrophys.~J., 131, 1142-1148. 
\\

\refs Stern, A.S.  2002.  Evidence for a collisional mechanism affecting the Kuiper Belt object colors.  Astron.~J., 124, 2297-2299.
\\

\refs Tegler, S.C. and Romanishin, W. 2000.  Extremely red Kuiper-belt objects in near-circular orbits beyond 40 AU.  Nature, 407, 979-981.
\\

\refs Tegler, S.C., Romanishin, W., Consolmagno, G.J.  2003.  Color patterns in the Kuiper Belt: A possible primordial origin.  Astrophys.~J., 599, L49-L52.
\\

\refs Th\'ebault, P. and Doressoundiram, A.  2003.  Colors and collision rates within the Kuiper Belt: Problems with the collisional resurfacing scenario.  Icarus, 162, 27-37.
\\

\refs Trujillo, C.A., Brown, M.E.  2002.  A correlation between inclination and color in the Classical Kuiper Belt.  Astrophys.~J., 566, L125-L128.



%
%

%
\begin{deluxetable}{lcllllccc}
\tabletypesize{\tiny}
\tablecaption{Classical Transneptunian Objects Observed with HST}
\tablewidth{0pt}
\tablehead{                                  &               &\multispan3{heliocentric orbital elements} &                                                                          &              &                     &                    \\ [-3pt]
\phantom{(123456)} object      &  binary  &  $a_{\odot}$ (AU) &  $e_{\odot}$ &\ $i_{\odot}$ ($^\circ$) &\ $\bar{i}$ ($^\circ$)  &$H_{\rm V}$ & instrument & time (UT)  \\ [-10pt] }
\startdata
\baselineskip=6pt

\phantom{(123456)} 2000 YF$_{2}$                         	&    	& 45.959  & 0.070  & 2.005   	& 0.602 & 7.0   	& A  & 08-04-06 06:05  \\ [-3pt] 
\phantom{(123456)} 1998 WG$_{24}$                     	&    	& 45.734  & 0.128  & 2.225    	& 0.761 & 6.6   	& A  & 11-16-05 23:31  \\ [-3pt] 
\phantom{(123456)} 2002 CZ$_{224}$                    	&     	& 45.262  & 0.055  & 1.687   	& 0.794 & 6.6   	& A  & 01-11-06 08:13  \\ [-3pt]  
\phantom{(123456)} 2001 QP$_{297}$                    	&    	& 45.267  & 0.119  & 1.431   	& 0.913 & 6.5   	& A  & 08-30-05 16:25  \\ [-3pt] 
\phantom{(123456)} 2001 QY$_{297}$                	&\ck	& 43.671  & 0.081  & 1.548   	& 0.957 & 5.4   	& A  & 04-18-06 16:53  \\ [-3pt] 
\phantom{(123456)} 2003 YP$_{179}$                    	&     	& 44.320  & 0.083  & 0.949	& 0.961 & 7.7   	& A  & 12-21-06 07:54  \\ [-3pt] 
\phantom{(123456)} 1997 CT$_{29}$                      	&     	& 43.494  & 0.035  & 1.015    	& 0.981 & 6.6   	& N  & 02-09-03 07:03  \\ [-3pt] 
\phantom{(123456)} 1998 WY$_{24}$                      	&     	& 43.345  & 0.041  & 1.912    	& 1.024 & 6.7   	& N  & 11-28-02 09:27  \\ [-3pt] 
\phantom{(123456)} 2003 QA$_{91}$                      	&\ck	& 44.177  & 0.069  & 2.417   	& 1.027 & 5.3   	& A  & 06-19-06 07:51  \\ [-3pt] 
\phantom{(123456)} 2002 WL$_{21}$                     	&    	& 43.382  & 0.046  & 2.551    	& 1.045 & 8.1   	& A  & 12-06-06 11:21  \\ [-3pt] 
\phantom{(123456)} 2000 QC$_{226}$                   	&    	& 44.250  & 0.054  & 2.662    	& 1.047 & 6.6   	& A  & 08-31-05 08:09  \\ [-3pt] 
\ \ (19255) 1994 VK$_{8}$                                          	&    	& 42.621  & 0.035  & 1.487    	& 1.112 & 7.0   	& N  & 09-08-02 00:40  \\ [-3pt] 
\phantom{(123456)} 1998 KY$_{61}$                      	&    	& 44.604  & 0.040  & 2.050   	& 1.135 & 7.3   	& N  & 05-01-03 22:40  \\  [-3pt] 
\ \ (88268) 2001 KK$_{76}$                                        	&     	& 42.245  & 0.015  & 1.889   	& 1.189 & 6.3   	& A  & 05-02-06 08:31  \\ [-3pt] 
\phantom{(123456)} 2000 CE$_{105}$                   	&    	& 44.061  & 0.060  & 0.547    	& 1.230 & 6.8   	& N  & 03-06-03 20:04  \\ [-3pt] 
\phantom{(123456)} 2003 TJ$_{58}$                       	&\ck 	& 44.575  & 0.089  & 0.958  	& 1.308 & 7.8   	& A  & 11-22-06 05:15  \\ [-3pt] 
(134860) 2000 OJ$_{67}$                                          	&\ck	& 42.880  & 0.018  & 1.114   	& 1.330 & 6.0   	& N  & 06-25-03 16:55  \\ [-3pt] 
\phantom{(123456)} 2000 CL$_{104}$                    	&     	& 44.359  & 0.079  & 1.247    	& 1.355 & 6.2   	& N  & 06-03-03 10:19  \\ [-3pt] %
\phantom{(123456)} 1999 OE$_4$                           	&    	& 45.568  & 0.052  & 2.149    	& 1.387 & 7.0   	& N  & 08-11-02 02:21  \\ [-3pt] 
\phantom{(123456)} 2001 OK$_{108}$                   	&    	& 43.039  & 0.033  & 1.996    	& 1.407 & 7.4   	& N  & 04-28-03 21:01  \\ [-3pt] 
\phantom{(123456)} 2003 YJ$_{179}$$^\flat$	 	&    	& 43.796  & 0.083  & 1.450   	& 1.548 & 7.0   	& A  & 11-21-06 05:16  \\ [-3pt] 
\ \  (66652)  1999 RZ$_{253}$                                     	&\ck	& 44.018  & 0.090  & 0.563   	& 1.564 & 5.9   	& N  & 04-23-03 02:26  \\ [-3pt] 
\ \ (88267) 2001 KE$_{76}$                                        	&    	& 42.746  & 0.026  & 0.496  	& 1.675 & 6.7  	& A  & 07-26-05 13:12  \\ [-3pt] 
\phantom{(123456)} 2001 OZ$_{108}$                    	&    	& 43.405  & 0.051  & 2.099   	& 1.689 & 8.4   	& A  & 07-29-06 15:00  \\ [-3pt] 
\ \ (66452) 1999 OF$_4$                                             	&    	& 45.130  & 0.061  & 2.655   	& 1.800 & 6.9   	& N  & 10-06-02 09:06  \\ [-3pt] 
\ \ (48639) 1995 TL$_{8}$$^\ddagger$                  	&     	& 52.558  & 0.239  & 0.247  	& 1.882 & 5.4  	& N  & 11-09-02 05:33  \\ [-3pt] 
\phantom{(123456)} 2001 FL$_{185}$                    	&\ck	& 44.178  & 0.077  & 3.559    	& 2.054 & 7.0  	& A  & 01-24-06 09:43  \\ [-3pt] 
\phantom{(123456)} 2001 OG$_{109}$                   	&    	& 43.635  & 0.019  & 0.547   	& 2.102 & 8.2  	& N  & 05-04-03 22:43  \\ [-3pt] 
\phantom{(123456)} 2003 WU$_{188}$                  	&\ck	& 44.437  & 0.039  & 3.769    	& 2.104 & 5.8  	& A  & 12-20-06 04:50  \\ [-3pt] 
\phantom{(123456)} 1998 KG$_{62}$                      	&     	& 43.674  & 0.048  & 0.787   	& 2.117 & 6.6  	& N  & 02-02-06 17:36  \\ [-3pt] 
\ \ (80806)  2000 CM$_{105}$      	                         	&\ck	& 42.201  & 0.070  & 3.756    	& 2.159 & 6.3  	& A  & 11-05-05 22:11  \\ [-3pt] 
\ \ (45802) 2000 PV$_{29}$                                        	&     	& 43.562  & 0.010  & 1.175   	& 2.186 & 8.0  	& A  & 07-26-06 16:40  \\ [-3pt] 
\phantom{(123456)} 2000 CQ$_{114}$                   	&\ck	& 46.009  & 0.115  & 2.699    	& 2.212 & 6.6  	& N  & 06-06-03 00:04  \\ [-3pt] 
\phantom{(123456)} 1999 OA$_{4}$                         	&    	& 44.526  & 0.061  & 3.120    	& 2.256 & 7.9  	& A  & 07-24-06 16:44   \\ [-3pt] 
\phantom{(123456)} 1998 WX$_{31}$                     	&    	& 45.658  & 0.111  & 2.965    	& 2.378 & 6.6  	& N  & 09-10-02 02:23  \\ [-3pt]  
\phantom{(123456)} 1999 RC$_{215}$                    	&    	& 44.427  & 0.057  & 1.399    	& 2.415 & 6.9  	& N  & 11-11-02 07:06  \\ [-3pt] 
\phantom{(123456)} 2001 RZ$_{143}$                    	&\ck	& 44.279  & 0.069  & 2.124    	& 2.433 & 6.0  	& A  & 09-04-05 22:51  \\ [-3pt] 
\phantom{(123456)} 2000 OU$_{69}$                      	&     	& 43.115  & 0.048  & 4.427 	& 2.471 & 6.3  	& N  & 10-09-02 18:44  \\ [-3pt] 
\ \ (33001) 1997 CU$_{29}$                                        	&    	& 43.227  & 0.041  & 1.459   	& 2.495 & 6.6  	& N  & 10-06-02 00:27  \\ [-3pt] 
\phantom{(123456)} 1999 OJ$_4$                           	&\ck	& 38.217  & 0.028  & 3.996    	& 2.602 & 7.0  	& N  & 10-04-02 13:50  \\ [-3pt] %
\ \                                                                                	&    	& \ \ \ \ \ "  & \ \ \  "&  \ \ \ \  "	& \ \ \ \  " &\  "   & A	& 07-24-05 16:45  \\ [-3pt] 
\phantom{(123456)} 2001 XR$_{254}$                    	&\ck	& 43.319  & 0.023  & 1.234   	& 2.638 & 5.6  	& A  & 12-20-06 06:28  \\ [-3pt] %
(123509) 2000 WK$_{183}$                                       	&\ck	& 44.774  & 0.050  & 1.965    	& 2.722 & 6.4  	& A  & 11-24-05 08:45  \\ [-3pt] 
\phantom{(123456)} 1999 RT$_{214}$                    	&\ck	& 42.711  & 0.052  & 2.575    	& 2.727 & 7.8  	& A  & 07-25-06 05:36  \\ [-3pt] 
\phantom{(123456)} 2002 FX$_{36}$                     	&     	& 44.375  & 0.048  & 1.126  	& 2.782 & 6.2   	& A  & 01-28-06 09:38  \\ [-3pt] %
(119951) 2002 KX$_{14}$             	                        	&  	& 38.606 & 0.046  & 0.406     	& 2.817 & 4.4  	&$A$& 04-07-06 10:36  \\ [-3pt] %
(126719) 2002 CC$_{249}$                                       	&     	& 47.203  & 0.194  & 0.837  	& 2.845 & 6.7  	& A  & 02-02-06 17:36  \\ [-3pt] 
\phantom{(123456)} 2000 CN$_{105}$                   	&    	& 44.512  & 0.102  & 3.421   	& 2.969 & 5.0   	& A  & 01-19-06 12:55  \\ [-3pt] %
\phantom{(123456)} 2002 PQ$_{145}$                   	&    	& 43.579  & 0.054  & 3.080    	& 3.215 & 5.6   & A  & 06-13-06 03:03  \\ [-3pt] 
(131695) 2001 XS$_{254}$                                       	&    	& 37.519  & 0.060  & 4.254    	& 3.310 & 7.7   & N  & 04-25-03 08:56  \\ [-3pt] 
\phantom{(123456)} 1999 HJ$_{12}$                       	&     	& 42.816  & 0.049  & 4.542  	& 3.544 & 7.4   & N  & 09-10-02 01:22  \\ [-3pt] 
\ \ (35671) 1998 FL$_{185}$                                       	&    	& 38.206  & 0.042  & 4.591   	& 3.611 & 5.8   & A  & 08-04-06 11:52  \\ [-3pt] 
\ \ (79360) 1997 CS$_{29}$                                         	&\ck	& 43.632  & 0.008  & 2.256    	& 3.840 & 5.1   & N  & 10-22-02 23:05  \\ [-3pt] 
\ \                                                                                	&    	& \ \ \ \ \ "  & \ \ \  "&  \ \ \ \  "	& \ \ \ \  " &\  "  & A  & 11-30-05 12:01  \\ [-3pt] 
\phantom{(123456)} 1999 CJ$_{119}$                    	&    	& 45.234  & 0.064  & 3.209    	& 4.629 & 7.4   & N  & 02-01-03 10:06  \\ [-3pt] 
\phantom{(123456)} 2001 KH$_{76}$                      	&    	& 45.842  & 0.119  & 3.263   	& 4.998 & 6.6   & A  & 04-20-06 08:41  \\ [-3pt] 
\phantom{(123456)} 2001 XU$_{254}$                    	&    	& 43.771  & 0.078  & 6.496  	& 4.998 & 6.3   & N  & 04-26-03 00:57   \\ [-3pt] 
(138537) 2000 OK$_{67}$                                         	&    	& 46.691  & 0.144  & 4.880    	& 5.000 & 6.0   & N  & 10-05-02 13:56  \\ [-3pt] %
\phantom{(123456)} 2003 QR$_{91}$                     	&\ck	& 46.362  & 0.183  & 3.503    	& 5.399 & 6.2   & A  & 06-10-06 06:19  \\ [-3pt] 
(148780) 2001 UQ$_{18}$                                         	&\ck	& 44.549  & 0.058  & 5.201   	& 5.461 & 5.1   & A  & 08-06-06 06:02  \\ [-3pt] %
\phantom{(123456)} 2000 PD$_{30}$                      	&     	& 46.724  & 0.025  & 4.588  	& 5.745 & 6.9   & N  & 05-07-03 00:25  \\ [-3pt] 
\phantom{(123456)} 1996 KV$_{1}$                        	&     	& 45.534  & 0.118  & 8.068   	& 6.190 & 7.3   & N  & 10-03-02 10:17  \\ [-3pt] 
\phantom{(123456)} 1997 CV$_{29}$                      	&     	& 41.945  & 0.041  & 8.042    	& 6.378 & 7.4   & N  & 10-17-02 02:15   \\ [-3pt] 
\phantom{(123456)} 2003 QQ$_{91}$                     	&    	& 38.671  & 0.075  & 5.406	& 7.023 & 7.5   & A  & 08-05-06 07:11   \\ [-3pt] 
\phantom{(123456)} 2000 CP$_{104}$                   	&    	& 44.179  & 0.099  & 9.492   	& 8.115 & 6.7   & N  	& 06-20-03 08:09  \\ [-3pt] 
\phantom{(123456)} 1993 FW                                    	&     	& 43.991  & 0.053  & 7.732   	& 8.134 & 7.0   & N  	& 03-15-03 17:25   \\ [-3pt] 
\phantom{(123456)} 2003 LD$_{9}$                        	&    	& 47.212  & 0.176  & 6.977	& 8.470 & 6.9   & A  	& 06-03-06 06:20   \\ [-3pt] 
\ \ (50000) Quaoar                                                        	&\ck	& 43.402  & 0.034  & 7.992    	& 8.521 & 2.6   &$A$& 02-14-06 20:58  \\ [-3pt] %
\phantom{(123456)} 2002 CY$_{248}$                   	&    	& 46.456  & 0.140  & 7.009  	& 8.588 & 5.1   & A  	& 01-09-06 11:25   \\ [-3pt] %
\phantom{(123456)} 1996 TS$_{66}$                      	&     	& 43.956  & 0.127  & 7.353 	& 8.828 & 6.6   & N  	& 08-22-98 10:42   \\ [-3pt] %
(150642) 2001 CZ$_{31}$                                           	&    	& 45.380  & 0.120  & 10.209 	& 9.013 & 5.7   & A  	& 01-18-06 08:05  \\ [-3pt] %
\phantom{(123456)} 2000 CJ$_{105}$                     	&     	& 44.074 & 0.111  & 11.615	& 10.916 & 5.9      & A	& 01-16-06 06:39  \\ [-3pt] %
\ \ (19521) Chaos                                                          	&     	& 45.755 & 0.105  & 12.042  	& 11.011 & 4.9      & A	& 10-27-02 19:47  \\ [-3pt] %
(134568) 1999 RH$_{215}$                                       	&     	& 43.973 & 0.156  & 10.209	& 11.143 & 8.1      & A  	& 07-28-06 02:18  \\ [-3pt] %
\phantom{(123456)} 1999 RN$_{215}$                    	&     	& 43.502 & 0.068  & 12.396 	& 11.348 & 6.7      & A  	& 10-06-05 06:10  \\ [-3pt] 
\phantom{(123456)} 1999 CQ$_{133}$                    	&     	& 41.226 & 0.093  & 13.294	& 11.428 & 7.0      & A  	& 04-25-03 20:13  \\ [-3pt] 
\phantom{(123456)} 1998 FS$_{144}$                     	&     	& 41.526 & 0.022  & 9.872   	& 11.601 & 6.7      & N  	& 06-19-03 01:50  \\ [-3pt] 
\phantom{(123456)} 1997 RT$_{5}$                        	&     	& 41.637 & 0.018  & 12.692  	& 12.685 & 7.3      & A  	& 10-04-02 07:33  \\ [-3pt] 
\phantom{(123456)} 2002 GJ$_{32}$                       	&     	& 44.148 & 0.108  & 11.640	& 13.160 & 5.5      & A  	& 04-25-06 00:24  \\ [-3pt] 
\phantom{(123456)} 2001 KA$_{77}$                  	&     	& 47.152 & 0.095  & 11.963	& 13.219 & 5.3      & A  	& 02-05-06 06:47  \\ [-3pt] 
\phantom{(123456)} 2003 YV$_{179}$$^\ddagger$	&    	& 47.486 & 0.227  & 15.538	& 14.211 & 7.2      & A  	& 12-23-06 07:52  \\ [-3pt] 
(118379) 1999 HC$_{12}$   $^\dagger$                 	&     	& 45.008 & 0.231  & 15.380   	& 14.559 & 7.6      & N  	& 05-14-03 11:15  \\ [-3pt] 
\ \ (20000) Varuna                                                      	&    	& 43.028 & 0.054  & 17.158	& 15.439 & 3.9      &{\sf A} 	& 11-05-05 19:01  \\ [-3pt] %
\ \                                                                                  	&    	& \ \ \ \ \ "  & \ \ \  "&  \ \ \ \  "	& \ \ \ \  " &\  "       &$A$	& 11-26-05 20:38  \\ [-3pt] %
\phantom{(123456)} 1999 OD$_{4}$  $^\dagger$	         &    	& 41.540 & 0.094  & 14.390	& 15.769 & 7.3      & N  	& 10-15-02 06:07  \\ [-3pt] 
\phantom{(123456)} 2002 CX$_{224}$                   	&     	& 46.498 & 0.131  & 16.820 	& 16.018 & 6.0      & A 	& 07-30-06 06:11  \\ [-3pt] 
\ \ (15883) 1997 CR$_{29}$      $^\dagger$            	&     	& 46.714 & 0.209  & 19.208  	& 17.854 & 7.2 	& N	& 10-18-02 20:09  \\ [-3pt] 
(143991) 2003 YO$_{179}$ $^\dagger$                 	&    	& 44.788 & 0.141  & 19.380	& 18.100 & 5.8 	& A	& 08-05-06 06:04  \\ [-3pt] %
\phantom{(123456)} 1999 CH$_{119}$ $^\dagger$	&    	& 43.234 & 0.084  & 19.987	& 18.413 & 7.3	& N  	& 10-24-02 02:26  \\ [-3pt] 
\ \ (55637) 2002 UX$_{25}$   $^\dagger$                 	&\ck	& 42.620 & 0.144  & 19.457  	& 20.232 & 3.6	& $A$ 	& 08-26-05 14:51  \\ [-3pt] %
\phantom{(123456)} 2000 CO$_{105}$ $^\dagger$	&    	& 47.190 & 0.150  & 19.266	& 20.760  & 5.4	& N 	& 02-25-03 04:01  \\ [-3pt] %
\phantom{(123456)} 1999 CL$_{119}$$^\dagger$	&    	& 46.793 & 0.003  & 23.328	& 21.814 & 5.9	& N    	& 10-22-02 21:31  \\ [-3pt] %
\ \ (86177) 1999 RY$_{215}$  $^\dagger$                 	&    	& 45.590 & 0.242  & 22.143  	& 23.307 & 7.1	& N   	& 04-28-03 22:44  \\ [-3pt] 
(120347) 2004 SB$_{60}$ $^\dagger$             		&\ck	& 42.068 & 0.110  & 23.903   	& 25.573 & 4.4	& A   	& 07-21-06 21:36  \\ [-3pt] %
\ \ (86047) 1999 OY$_{3}$ $^\dagger$                      	&    	& 43.933 & 0.169  & 24.219  	& 25.818 & 7.1	& A    	& 08-02-06 10:07  \\ [-3pt] 
\ \ (55565) 2002 AW$_{197}$  $^\dagger$             	&    	& 46.965 & 0.128  & 24.449  	& 26.009 & 3.6	&$A$ 	& 12-01-05 10:50  \\ [-3pt] 
\phantom{(123456)} 1999 OH$_{4}$$^\dagger$   	&    	& 40.571 & 0.041  & 28.166	& 26.699 & 8.3	& A   	& 08-03-06 10:07  \\ [-3pt] 
\ \ (24835) 1995 SM$_{55}$  $^\dagger$                    	&    	& 41.641 & 0.102  & 27.116  	& 26.793 & 4.8	& N   	& 10-02-02 02:47  \\ [-3pt] %
\ \ (55636) 2002 TX$_{300}$  $^\dagger$                   	&    	& 43.086 & 0.120  & 25.901  	& 26.976 & 3.3	&$A$ 	& 09-16-05 08:27  \\ [-3pt] %
\ \ (19308) 1996 TO$_{66}$ $^\dagger$                      	&    	& 43.371 & 0.122  & 27.439  	& 27.954 & 4.5	& N   	& 08-21-98 10:42  \\ [-3pt] %
\ \                                                                                   	&    	& \ \ \ \ \ "  & \ \ \  "&  \ \ \ \  "	& \ \ \ \  " &\  "   &$A$	& 07-11-05 23:19  \\ [-3pt] %
\phantom{(123456)} 2000 CG$_{105}$$^\dagger$  	&    	& 46.362 & 0.043  & 28.001	& 29.379 & 6.5	& N    	& 04-27-03 18:41  \\ [-3pt] 
\phantom{(123456)} 2001 QC$_{298}$$^\dagger$  	&\ck 	& 46.446 & 0.127  & 30.552	& 31.540 & 6.1	& N    	& 10-04-02 15:30  \\ [-3pt] 
\phantom{(123456)} 1996 RQ$_{20}$$^\dagger$  	&    	& 44.016 & 0.104  & 31.635	& 31.708 & 7.0	& N   	& 10-04-02 01:13  \\ [-3pt] 
(149560) 2003 QZ$_{91}$$^\ast$                          	&    	& 41.318 & 0.475  & 34.885	& 34.319 & 5.9	& A   	& 08-03-06 11:45  \\ [-3pt] 

\enddata

\tablecomments {Objects are listed in order of increasing $\bar {i}$, their mean inclination with respect to the invariable plane.  Mean inclinations, $\bar {i}$ are averaged over 10 Myr integrations (Elliot et al.~2005).  $^\ddagger$ Classified as Scattered-Extended by Elliot et al.~(2005); as Classical by Gladman et al.~(2008).  $^\flat$ Classified as 7:4EEE by the DES.  Both $\pm$1-sigma orbits are, however, Classical.  Classified as Classical by Gladman et al.~(2008).  $^\dagger$ Classified as Scattered-Near by Elliot et al.~(2005); as Classical by Gladman et al.~(2008).  $^\ast$ Classified  as a Centaur by Elliot et al.~(2005); as Classical by Gladman et al.~(2008).  A and N in the instruments column denote observations made by the authors with the ACS/HRC and NICMOS/NIC2 cameras respectively.  The first observation of Varuna was made by Sheppard and Noll in HST program 10555 and is marked by {\sf A}.  Objects denoted by an italic $A$ were observed by Brown and colleagues in HST program 10545 (Brown and Suer, 2007).  (55636) 2002 TX$_{300}$ was observed by both Brown et al.~and by Noll et al.~in program 10801. }

\end{deluxetable}

\begin{deluxetable}{lccccc}
\tabletypesize{\small}
\tablecaption{Statistical Tests}
\tablewidth{0pt}
\tablehead{     				&                 &                  	&  \multispan2{\ \  K-S }    		& Yates $\chi^2$   			\\ [1pt]
\ \ \ sample      				& singles	& binaries    	& $D$  &   $p_{\rm KS}$          	& $p_\chi$(5.5$^\circ$)        	\\ [-10pt] }
\startdata
\baselineskip=6pt

GMV,  $H_{\rm V} > 5$               	&  73	   &  18 	    &  0.423  	&  0.008		& 0.0045     				\\ \medskip
DES,  $H_{\rm V} > 5$           	   	&  57	   &  17 	    &  0.380  	&  0.033  		& 0.0098   				\\
GMV, all $H_{\rm V}$                 	&  80	   &  21 	    &  0.324  	&  0.047 		& 0.0277     				\\
DES , all $H_{\rm V}$            		&  60	   &  18 	    &  0.339  	&  0.064 		& 0.0327     				\\
\enddata

\tablecomments {$D$ is the largest vertical difference between the cumulative distribution functions of singles and binaries as functions of inclination.  $p_{\rm KS}$ is the probability of obtaining a given $D$ from random variation (Press et al.~1992). }

\end{deluxetable}

%
%

\begin{figure*} 
\epsscale{1.}
\plotone{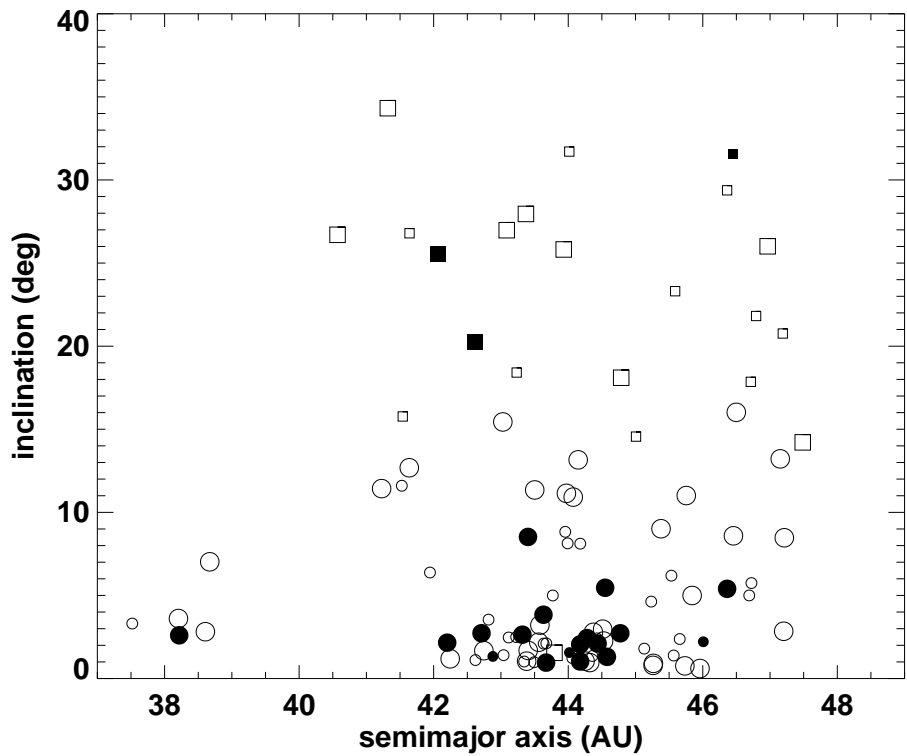}
\caption{\small  Classical TNOs are shown in this figure with their mean orbital inclination realtive to the invariable plane, $\bar {i}$, plotted against semimajor axis, $a$.  Binaries are shown as filled symbols; apparently single objects are denoted by open symbols.  Objects shown with circular symbols are considered Classicals by both the Elliot et al.~(2005) and Gladman et al.~(2008).  Objects represented with square symbols are considered as Classicals by Gladman et al., but not by Elliot et al..  Larger symbols are objects observed with the HRC (0.025 arcsec pixel scale).  Smaller symbols are objects observed with NIC2 (0.075 arcsec scale).   One object, (48639) 1995 TL$_8$ identified as Classical by Gladman et al.~lies off the plot to the right with $a =$ 52.6 AU and $\bar {i}= 1.9 ^\circ$.  The strong anticorrelation of binaries with inclination is obvious.}  
\end{figure*}

\begin{figure*} 
\epsscale{1.}
\plotone{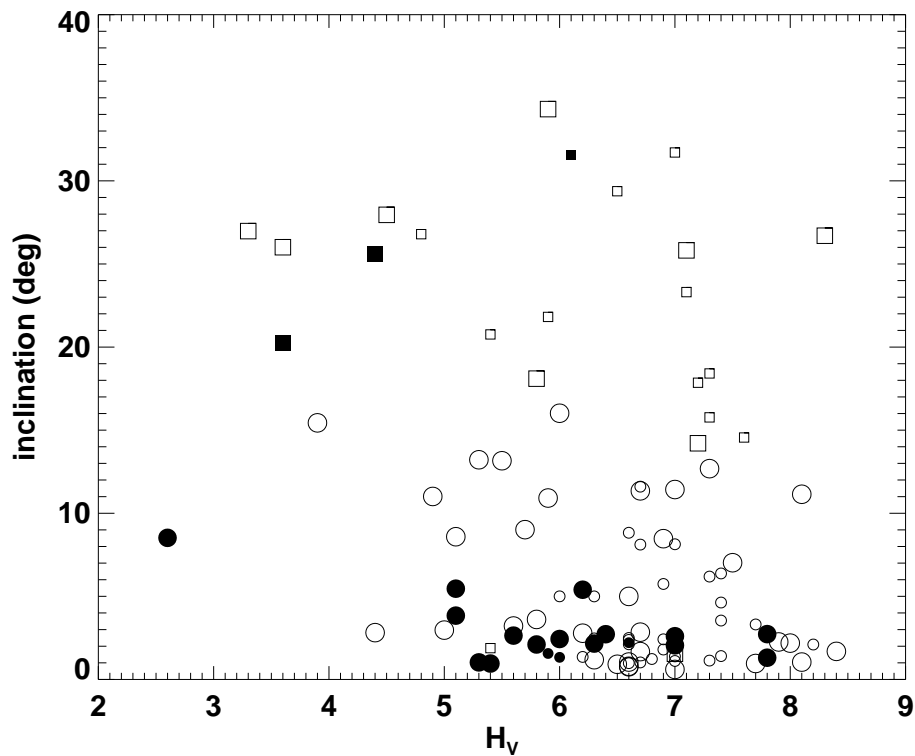}
\caption{\small  Classical TNOs mean orbital inclination, $\bar {i}$, is plotted against absolute magnitude, $H_{\rm V}$.  Symbols have the same meaning as in Figure 1.   Only one of the high inclination binaries is a similar-mass binary like the low-inclination systems.  The three binaries with $H_{\rm V} < 5 $ are the only systems in this sample with companions that are significantly fainter than their primary ($\Delta_{\rm mag} > 2$; Noll et al.~2008). }  
\end{figure*}

\begin{figure*}
\epsscale{1.}
\plotone{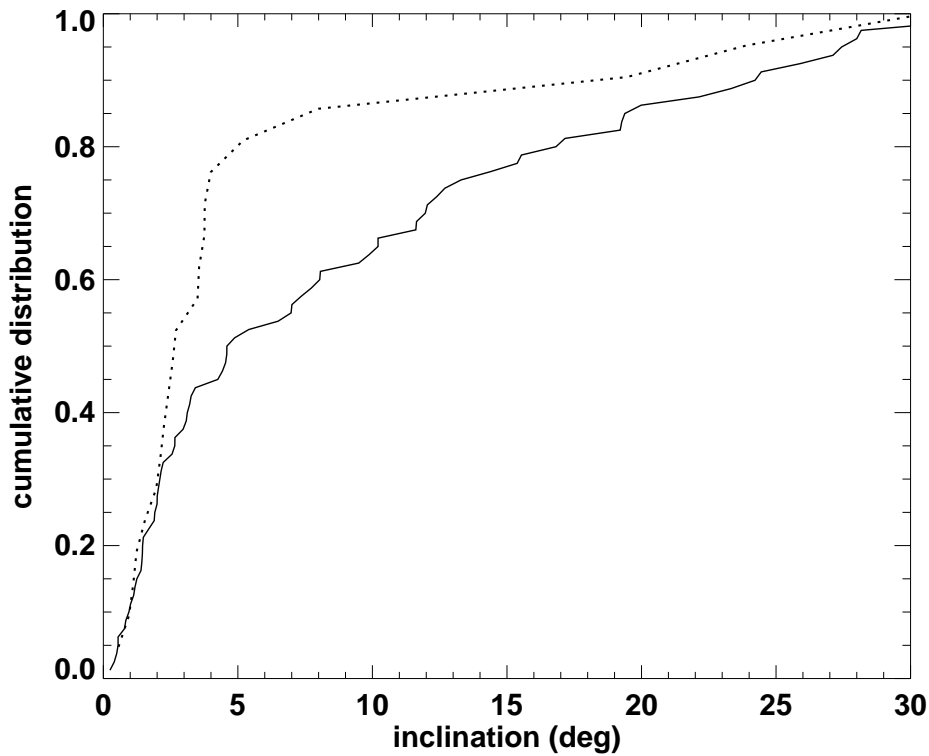}
\caption{\small  The cumulative distribution (CDF) of apparently single TNOs in our sample is plotted against ($\bar {i}$) as a solid curve.  The CDF of binaries is shown by a dotted curve.  Our sample mimics the previously reported distribution of TNOs with inclination showing a concentration at low $\bar {i}$.  The CDF for binaries is much more strongly concentrated to low inclinations.  The few high inclination systems plotted here include the small satellites of very large TNOs which are arguably distinct from the similar-mass binaries that are ubiquitous at small inclinations.  The likelihood of these two distributions arising by chance from a common parent population is less than 5\% (see Table 2). If we restrict the sample to objects with absolute magnitudes $H_{\rm V} > 5 $ the probability is less than 1\%.}  
\end{figure*}

\begin{figure*} 
\epsscale{1.}
\plotone{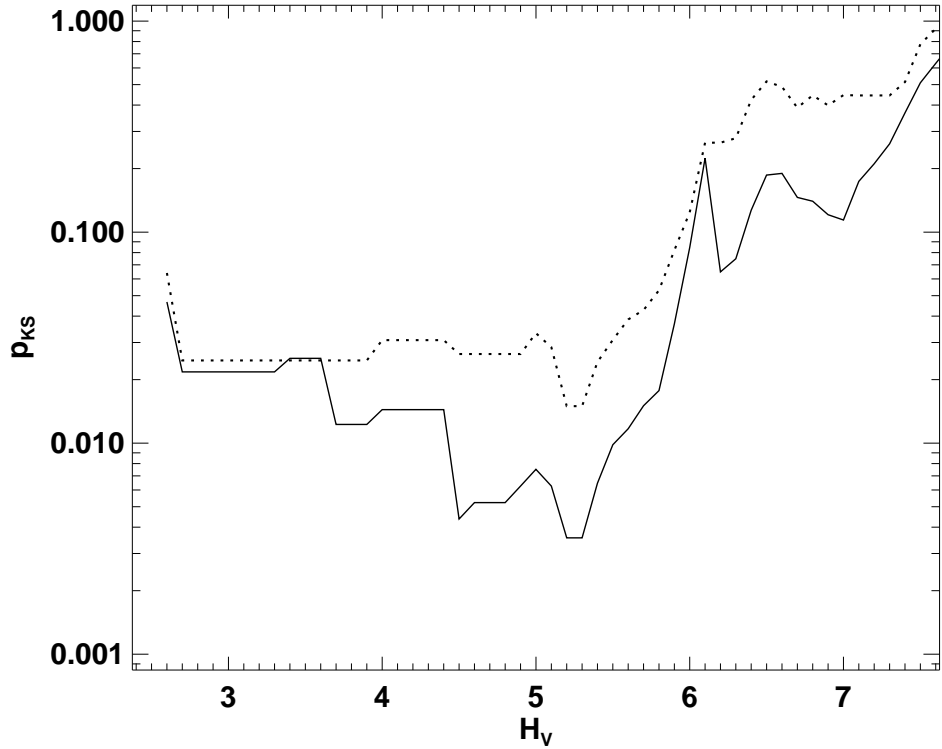}
\caption{\small  The probability, $p_{\rm KS}$, that the cumulative distribution function (CDF, see Fig.~3) of binaries and singles come from a single population is plotted as a function of the minimum absolute magnitude of the sample.  A broad minimum in this probability occurs near $H_{\rm V}$ = 5.  Curves are shown both for the DES-defined Classicals (dotted) and for the more inclusive GMV definition (solid). }  
\end{figure*}

\end{document}